\documentclass[aps,prd,twocolumn,superscriptaddress,showpacs,preprintnumbers,amsmath,amssymb]{revtex4}
\usepackage{epsfig}
\usepackage{rotating}
\usepackage{graphicx}               
\usepackage{pslatex}               
\usepackage{dcolumn}                
\usepackage{multirow}               
\usepackage{amsmath}                
\usepackage{color}
\usepackage{comment}
\graphicspath{{ps}}                 

\renewcommand{\arraystretch}{1.1}   

\newcommand{\plpi}{{p\overline{\Lambda}\pi^-}}

\newcommand{\pld} {\overline{p} \Lambda D^0}
\newcommand{\pldst} {\overline{p} \Lambda D^{*0}}
\newcommand{\pldstp} {\overline{p} \Lambda D^{*+}}
\newcommand{\pldstzp} {\overline{p} \Lambda D^{(*)0}}
\newcommand{\lr} {\mathcal R}
\newcommand{\plbar}{\overline{p}{\Lambda}}

\newcommand{\lam}{\Lambda}

\newcommand{\pldtwoyd}{26.5^{+6.3}_{-5.6}}
\newcommand{\pldthryd}{35.6^{+11.7}_{-10.7}}
\newcommand{\pldstzyd}{4.3^{+3.2}_{-2.4}}

\newcommand{\pldtwobf}{(1.39^{+0.33}_{-0.29} \pm 0.16)\times 10^{-5}}
\newcommand{\pldthrbf}{(1.54^{+0.50}_{-0.46} \pm 0.26)\times 10^{-5}}
\newcommand{\pldtotbf}{(1.43^{+0.28}_{-0.25} \pm 0.18)\times 10^{-5}}
\newcommand{\pldstzbf}{(1.53^{+1.12}_{-0.85} \pm 0.47)\times 10^{-5}}

\newcommand{\pldtwobfx}{1.39^{+0.33}_{-0.29} \pm 0.16}
\newcommand{\pldthrbfx}{1.54^{+0.50}_{-0.46} \pm 0.26}
\newcommand{\pldtotbfx}{1.43^{+0.28}_{-0.25} \pm 0.18}
\newcommand{\pldstzbfx}{1.53^{+1.12}_{-0.85} \pm 0.47}

\newcommand{\mpl}{{M_{\overline{p}\Lambda}}}
\newcommand{\dm}{\Delta M}

\newcommand{\mb}{M_{\rm bc}}
\newcommand{\de}{{\Delta{E}}}
\newcommand{\bp}{B^{+}}
\newcommand{\bz}{B^{0}}
\newcommand{\bm}{B^{-}}

\begin{document}
\preprint{\vbox{ 
		    \hbox{Belle Preprint  2011-9}
	               \hbox{KEK Preprint  2011-7}
}}

\title{ \quad\\[0.5cm] Observation of $\bm \to \pld$ at Belle}


\affiliation{Budker Institute of Nuclear Physics SB RAS and Novosibirsk State University, Novosibirsk 630090}
\affiliation{Faculty of Mathematics and Physics, Charles University, Prague}
\affiliation{University of Cincinnati, Cincinnati, Ohio 45221}
\affiliation{Department of Physics, Fu Jen Catholic University, Taipei}
\affiliation{Justus-Liebig-Universit\"at Gie\ss{}en, Gie\ss{}en}
\affiliation{Gifu University, Gifu}
\affiliation{Hanyang University, Seoul}
\affiliation{University of Hawaii, Honolulu, Hawaii 96822}
\affiliation{High Energy Accelerator Research Organization (KEK), Tsukuba}
\affiliation{Hiroshima Institute of Technology, Hiroshima}
\affiliation{Indian Institute of Technology Guwahati, Guwahati}
\affiliation{Indian Institute of Technology Madras, Madras}
\affiliation{Institute of High Energy Physics, Chinese Academy of Sciences, Beijing}
\affiliation{Institute of High Energy Physics, Vienna}
\affiliation{Institute of High Energy Physics, Protvino}
\affiliation{Institute for Theoretical and Experimental Physics, Moscow}
\affiliation{J. Stefan Institute, Ljubljana}
\affiliation{Kanagawa University, Yokohama}
\affiliation{Institut f\"ur Experimentelle Kernphysik, Karlsruher Institut f\"ur Technologie, Karlsruhe}
\affiliation{Korea Institute of Science and Technology Information, Daejeon}
\affiliation{Korea University, Seoul}
\affiliation{Kyungpook National University, Taegu}
\affiliation{\'Ecole Polytechnique F\'ed\'erale de Lausanne (EPFL), Lausanne}
\affiliation{Faculty of Mathematics and Physics, University of Ljubljana, Ljubljana}
\affiliation{Luther College, Decorah, Iowa 52101}
\affiliation{University of Maribor, Maribor}
\affiliation{Max-Planck-Institut f\"ur Physik, M\"unchen}
\affiliation{University of Melbourne, School of Physics, Victoria 3010}
\affiliation{Nagoya University, Nagoya}
\affiliation{Nara Women's University, Nara}
\affiliation{National Central University, Chung-li}
\affiliation{National United University, Miao Li}
\affiliation{Department of Physics, National Taiwan University, Taipei}
\affiliation{H. Niewodniczanski Institute of Nuclear Physics, Krakow}
\affiliation{Nippon Dental University, Niigata}
\affiliation{Niigata University, Niigata}
\affiliation{University of Nova Gorica, Nova Gorica}
\affiliation{Osaka City University, Osaka}
\affiliation{Pacific Northwest National Laboratory, Richland, Washington 99352}
\affiliation{Panjab University, Chandigarh}
\affiliation{Research Center for Nuclear Physics, Osaka}
\affiliation{University of Science and Technology of China, Hefei}
\affiliation{Seoul National University, Seoul}
\affiliation{Sungkyunkwan University, Suwon}
\affiliation{School of Physics, University of Sydney, NSW 2006}
\affiliation{Tata Institute of Fundamental Research, Mumbai}
\affiliation{Excellence Cluster Universe, Technische Universit\"at M\"unchen, Garching}
\affiliation{Toho University, Funabashi}
\affiliation{Tohoku Gakuin University, Tagajo}
\affiliation{Tohoku University, Sendai}
\affiliation{Department of Physics, University of Tokyo, Tokyo}
\affiliation{Tokyo Institute of Technology, Tokyo}
\affiliation{Tokyo Metropolitan University, Tokyo}
\affiliation{Tokyo University of Agriculture and Technology, Tokyo}
\affiliation{CNP, Virginia Polytechnic Institute and State University, Blacksburg, Virginia 24061}
\affiliation{Yonsei University, Seoul}

 \author{P.~Chen}\affiliation{Department of Physics, National Taiwan University, Taipei} 
 \author{M.-Z.~Wang}\affiliation{Department of Physics, National Taiwan University, Taipei} 
  \author{I.~Adachi}\affiliation{High Energy Accelerator Research Organization (KEK), Tsukuba} 
  \author{H.~Aihara}\affiliation{Department of Physics, University of Tokyo, Tokyo} 
  \author{D.~M.~Asner}\affiliation{Pacific Northwest National Laboratory, Richland, Washington 99352} 
  \author{V.~Aulchenko}\affiliation{Budker Institute of Nuclear Physics SB RAS and Novosibirsk State University, Novosibirsk 630090} 
  \author{T.~Aushev}\affiliation{Institute for Theoretical and Experimental Physics, Moscow} 
  \author{A.~M.~Bakich}\affiliation{School of Physics, University of Sydney, NSW 2006} 
  \author{E.~Barberio}\affiliation{University of Melbourne, School of Physics, Victoria 3010} 
  \author{K.~Belous}\affiliation{Institute of High Energy Physics, Protvino} 
% ...
\collaboration{The Belle Collaboration}

\noaffiliation

 \begin{abstract} 

We study $\bm$ meson decays to $\overline{p} \Lambda D^{(*)0}$ final states  using  a sample of $657 \times 10^6~B\overline{B}$ events
collected at the $\Upsilon(4S)$ resonance with the Belle detector at the 
KEKB asymmetric-energy $e^+ e^-$ collider. 
The observed branching fraction for $\bm \to \pld$ is 
$\pldtotbf$ with a significance of 8.1 standard deviations, where the uncertainties are statistical and systematic, respectively.
Most of the signal events have the $\plbar$ mass peaking near threshold. 
No significant signal is observed for $\bm \to \pldst$ and the corresponding upper limit on the branching fraction is $4.8 \times 10^{-5}$ at the 90\% confidence level.

 \end{abstract}  
\maketitle

\renewcommand{\thefootnote}{\fnsymbol{footnote}}
\setcounter{footnote}{0}
\clearpage

 Since the first observations of baryonic decays of $B$ mesons
by ARGUS~\cite{1} and CLEO~\cite{2}, many three-body baryonic $B$ 
decays have been found~\cite{3}. 
Although the general pattern of these decays 
can be understood intuitively from heavy $b$ quark decays~\cite{4}, 
many specific details cannot be explained by this simple picture. 

Using a generalized factorization 
approach, Ref.~\cite{5} predicts rather large branching fractions 
($\sim$$10^{-5}$) for the Cabibbo-suppressed processes
$B \to \overline{p} \Lambda  D^{(*)} $. 
The branching fractions of other related 
baryonic decays such as $B^0 \to p\overline{p}D^0$~\cite{6,7},
~$B^0 \to p\overline{p}K^{*0}$~\cite{8}, $B^- \to p\overline{p}K^{*-}$~\cite{9,10} and 
$B^- \to p\overline{p}\pi^-$~\cite{9} are used as
inputs in such estimates because  baryon form factors entering
the decay amplitudes are difficult to calculate from first principles. 
The expected values of the branching fractions for  $B^- \to \overline{p}\Lambda D^0$ and 
$B^- \to \overline{p}\Lambda D^{*0}$
are already within reach with the data sample accumulated at Belle.

Nearly all  baryonic $B$ decays into three- and
four-body final states possess a common feature: baryon-antibaryon 
invariant masses that peak near threshold. This 
threshold enhancement is found both in charmed and charmless 
cases~\cite{3}. A similar effect has been observed 
in $J/\psi \to p\overline{p}\gamma$ decays by BES~\cite{bes1,bes2} 
and CLEO~\cite{cleo1}, but is not seen in 
$J/\psi \to p \overline{p} \pi^0 $~\cite{bes1} and 
$\Upsilon(1S) \to  p\overline{p}\gamma$~\cite{cleo2}. One of the possible 
explanations of this phenomenon suggested in the literature is a final state 
$N\overline{N}$ interaction~\cite{int}.

In this paper, we present results on the $B^- \to \overline{p}\Lambda D^{(*)0}$ decays in order to test the  factorization hypothesis
and study the $\overline{p}\Lambda$ threshold enhancement effect.


The data sample used in the study corresponds to an integrated luminosity of 605 
fb$^{-1}$, containing 657 $ \times 10^6~B\overline{B}$ pairs, collected at the $\Upsilon(4S)$ resonance
with the Belle detector at the KEKB asymmetric-energy
$e^+e^-$ (3.5~GeV and 8~GeV) collider~\cite{KEKB}.
The Belle detector~\cite{Belle} is a large-solid-angle magnetic spectrometer that consists of
 a silicon vertex detector (SVD),
a 50-layer central drift chamber (CDC), an array of aerogel threshold Cherenkov
counters (ACC), a barrel-like arrangement of time-of-flight scintillation counters 
(TOF), and an electromagnetic calorimeter (ECL) composed of CsI(Tl) crystals 
located inside a superconducting solenoid coil that provides a 1.5~T magnetic field.

The selection criteria for the final state charged particles in $\bm \to \pld$ and $\bm \to \pldst$ are 
based on information obtained from the tracking system (SVD and CDC) and 
the hadron identification system (CDC, ACC, and TOF). 
The primary and $D^0$ daughter charged tracks
are required to have a point of
closest approach to the interaction point (IP) that is within
$\pm$0.3 cm in the transverse ($x$--$y$) plane, and within $\pm$3.0 cm
in the $z$ direction, where the $+z$ axis is opposite to the
positron beam direction.
For each track, the likelihood values $L_p$, $L_K$, or $L_\pi$ that it is a proton, kaon, or pion,
respectively, are determined from the information provided by the hadron identification system. 
A track is identified as a proton if $L_{p}/(L_p +L_{K}) > 0.6$ and $L_{p}/(L_p +L_{\pi}) > 0.6$, as a kaon
if $L_{K}/(L_K +L_{\pi}) > 0.6$, or as a pion if $L_{\pi}/(L_K +L_{\pi}) > 0.6$.
 The efficiency for identifying a kaon (pion) is 85$-$95\% depending on the momentum of the track, while the probability for a pion
(kaon) to be misidentified as a kaon (pion) is 10$-$20\%.
The proton identification efficiency is 84\% while the probability for a kaon or a pion to be misidentified as a proton is less than 10\%.

We reconstruct $\lam$'s from their decays to $p \pi^-$. 
Each $\Lambda$ candidate must have a displaced vertex and the direction of its momentum vector must be
consistent with an origin at the IP.
The proton-like daughter is required to satisfy the proton criteria described above, and no further selections are applied
to the daughter tracks. 
The reconstructed $\lam$ mass is required to be in the range 1.111 GeV/$c^2 < M_{p\pi^-} <1.121 $ GeV/$c^2$~\cite{3}.

Candidate $D^0$ mesons are reconstructed in the following two sub-decay channels: 
$D^0 \to K^-\pi^+$ and $D^0 \to K^-\pi^+\pi^0$, $\pi^0 \to \gamma\gamma$. 
The $\gamma$'s that constitute $\pi^0$ candidates are required to have 
energies greater than 50 MeV if the $\gamma$ is reconstructed from the barrel ECL
and greater than 100 MeV for the endcap ECL, and not be associated with any charged
tracks in CDC.
The energy asymmetry of $\gamma$'s from a $\pi^0 $, $\frac{|E_{\gamma1}-E_{\gamma2}|}{E_{\gamma1}+E_{\gamma2}}$, 
is required to be less than 0.9.
The mass of a $\pi^0$ candidate is required to be within the range 
$0.118$ GeV/$c^2 < M_{\gamma\gamma}<0.150$ GeV/$c^2$ before a mass-constrained fit is applied to improve the $\pi^0$ momentum resolution. We impose a cut on the invariant masses of the $D^0$ candidates, $|M_{K^-\pi^+}-1.865$ GeV/$c^2|<0.01$ GeV/$c^2$ and $1.837$ 
GeV/$c^2 <M_{K^-\pi^+\pi^0}<1.885$ GeV/$c^2$ for $D^0 \to K^-\pi^+$ and $D^0 \to K^-\pi^+\pi^0$, respectively, which retains about 87\% of the signal.

We reconstruct $D^{*0}$ mesons in the decay mode $D^{*0} \to D^0 \pi^0$ 
with $D^0 \to K^- \pi^+$ only. 
Since the $\pi^0$ coming from the $D^{*0}$ decay is expected to have low energy, we adjust the
photon selection criteria accordingly.
The energy of $\gamma$'s that constitute $\pi^0$ candidates from a $D^{*0}$ must be greater than 50 MeV.
The energy asymmetry of the two $\gamma$'s is required to be less than 0.6 
and the di-photon invariant mass should be in the range $0.120$ GeV/$c^2< M_{\gamma\gamma}<0.158$ GeV/$c^2$. 
For the $D^{*0}$ candidates, we require $0.139$ GeV/$c^2< \dm < 0.145$ GeV/$c^2$, where $\dm$ denotes the mass difference between $D^{*0}$ and $D^{0}$. 

Candidate $B$ mesons are identified with two kinematic variables calculated 
in the center-of-mass (CM) frame: the beam-energy-constrained mass 
$\mb = \sqrt{E^2_{\rm beam}-p^2_B}$, and the energy difference $\de = E_B - E_{\rm beam}$, 
where $E_{\rm beam}$ is the beam energy, and $p_B$ and $E_B$ are the momentum and energy, 
respectively, of the reconstructed $B$ meson. 
In order to reduce the contribution from combinatoric backgrounds, we define the candidate region for $\bm \to \pld$ ($\pldst$) as $5.2$ GeV/$c^2 < \mb < 5.3$ GeV/$c^2$, $-0.1$ GeV $ < \de< 0.4$ GeV and $\mpl< 3.4$ ($3.3$) GeV/$c^2$, where $\mpl$ denotes the invariant mass of the baryon pair.
The lower bound in $\de$ is chosen to exclude backgrounds from multibody baryonic $B$ decays.
From Monte Carlo (MC) simulations based on GEANT~\cite{geant},
we define the signal region as $5.27$ GeV/$c^2 < \mb < 5.29$ GeV/$c^2$ 
and $|\de|< 0.05$ GeV.

The dominant background for $\bm \to \pld$ in the candidate region is from continuum $e^+e^-
\to q\overline{q}$ ($q = u,\ d,\ s,\ c$) processes. 
We suppress the jet-like continuum background relative to the more
spherical $B\overline{B}$ signal using a Fisher discriminant 
that combines seven event-shape variables derived from modified Fox-Wolfram moments~\cite{rr-104} as described in Ref.~\cite{rr-551}.
The Fisher discriminant is a linear combination of several variables with coefficients that are optimized to separate signal and background.
In addition to the Fisher discriminant, two variables $\cos\theta_{B}$ and $\Delta z$ are used to form signal and background probability density functions (PDFs). 
The variable $\theta_{B}$ is the angle between the reconstructed $B$ direction and
the beam axis in the CM frame,
and $\Delta z$ is the difference between the $z$ positions of the candidate $B$ vertex 
and the vertex of the rest of the final state particles, presumably, from the other $B$ in the $\Upsilon({\rm 4S})$ decay. 
The products of the above PDFs, obtained from signal and continuum MC simulations, 
give the event-by-event signal and background likelihoods, ${\mathcal L}_S$ and ${\mathcal L}_B$. 
We apply a selection on the likelihood ratio, ${\lr} = {\mathcal L}_S/({\mathcal L}_S+{\mathcal L}_B)$ to suppress background.
Information associated with the accompanying B meson can also be used to distinguish $B$ events from continuum events. The
variables used are ``$q$'' and ``$r$'' from a $B$ flavor-tagging algorithm~\cite{flavor_tag}. 
The value of the preferred flavor $q$ equals $+1$ for $B^0 / B^+$ 
and $-1$ for $\overline{B}^0 / B^-$. 
The $B$ tagging quality factor $r$ ranges from zero
for no flavor information to unity for unambiguous flavor assignment.
Sets of $q \times r$-dependent $\lr$  selection requirements are optimized by maximizing a figure of merit defined as $N_S / \sqrt{N_S+N_B}$, where $N_{S}$ denotes the expected number of signal events based on MC simulation and the predicted branching fraction, and $N_{B}$ denotes the expected number of background events from the continuum MC.
The requirements on $\lr$ remove 75\% (89\%) of continuum background
while retaining 88\% (69\%) of the signal for $\bm \to \pld$ with $D^0 \to K^-\pi^+$ ($ D^0 \to K^-\pi^+\pi^0$). 
The continuum background suppression is not applied for $\bm \to \pldst$ since the optimal $\lr$ requirement is close to zero.

\begin{figure}[htb]
\vskip -3.1cm
\begin{sideways}  ~~~~~~~~~~~~~~~~~~~~~~Events / (0.8 MeV/$c^2$)  \end{sideways}
\hskip -0.5cm
\includegraphics[width=0.32\textwidth]{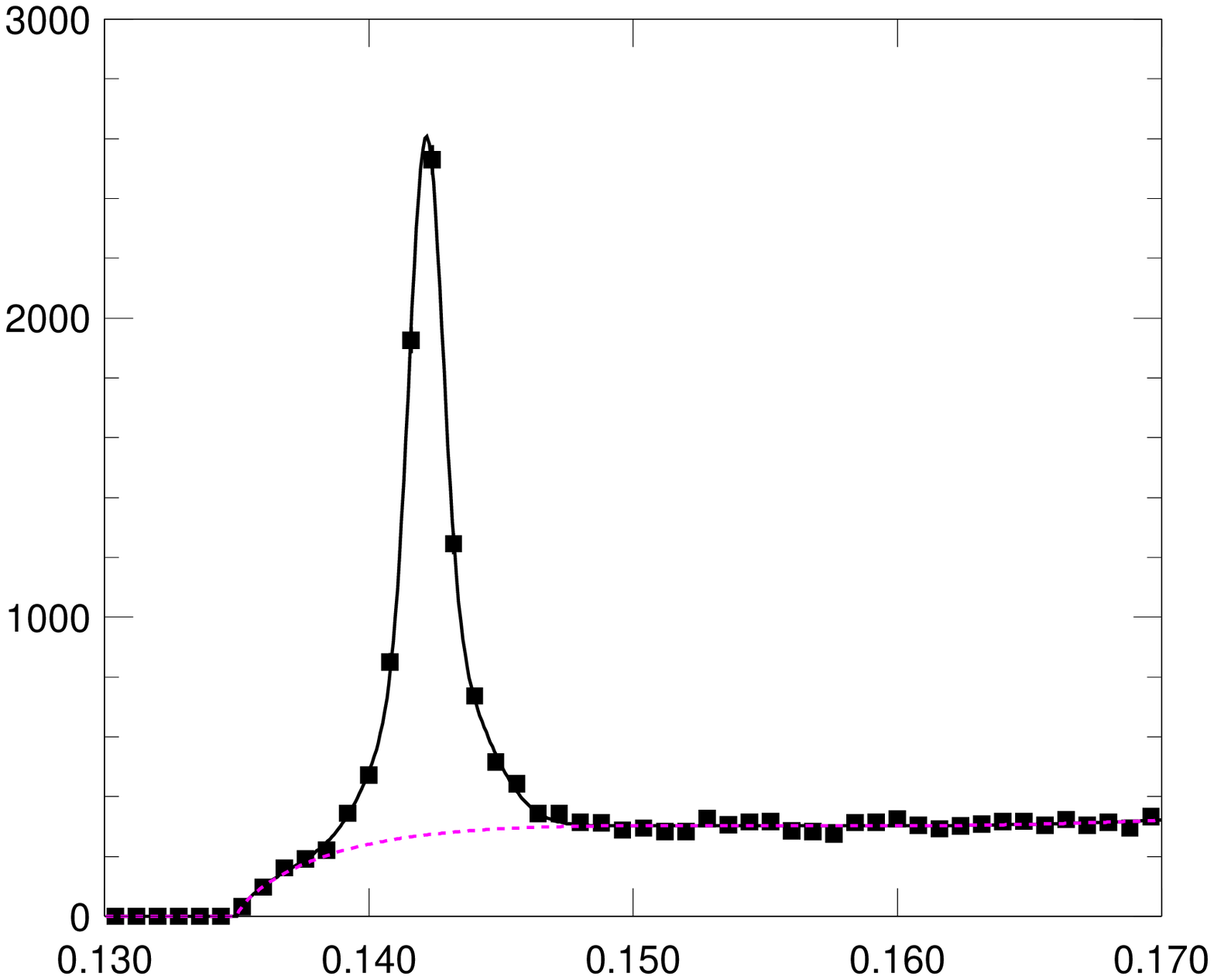}\\
\vskip -0.3cm
\vskip -4.cm \hskip 4.cm{MC}
\vskip 3.6cm 
\hskip 3.5cm {$\Delta M $ (GeV/$c^2$)}
\caption{
The $\dm$ distribution of the $\bm \to \pldst$ MC sample with fit curves overlaid,  where $\dm$ denotes the mass difference between $D^{*0}$ and $D^{0}$. 
The solid curve is the overall fit result, the dashed curve shows the CF background and the black filled squares are the MC events. }
\label{fg:fit_dm_scf3}
\end{figure}

In order to avoid multiple counting, 
in cases where more than one $B$ candidate is found in a single event,
we choose the one with the smallest $\chi^2 = \chi^2_{B} + \chi^2_{\lam} (+\chi^2_{\pi^0})$, where $\chi^2$ 
is calculated from the vertex fit to the $B$ using $\overline{p}$ and $D^{0}$ measurements, the vertex fit to $\Lambda$ using $p$ and $\pi^-$ tracks, and the mass-constrained $\pi^0$ fit if applicable. 
The fraction of events with multiple candidates are 2.3\% (14.1\%) of the sample according to 
MC simulations for $\bm \to \pld$ with $D^0 \to K^-\pi^+$ ($D^0 \to K^-\pi^+ \pi^0$), 
and 17.8\% for $\bm \to \pldst$. 
The dominant background for $\bm \to \pldst$ is 
from $\overline{ B^0} \to \pldstp$ cross-feed and $\bm \to \pldst$ self cross-feed (both referred to as CF) events according to a MC simulation based on \textsc{Pythia}~\cite{Pythia}.
In CF events,  two low-energy $\gamma$'s can form a $\pi^0$ candidate that is combined with a correctly reconstructed  $D^0$, $\overline{p}$
and $\Lambda$ from a $B$ decay to form a candidate event in the signal region.
These backgrounds cannot be distinguished from the signal in the $\mb - \de$ two-dimensional fit alone, 
although their distributions in $\mb$ and $\de$ have a slightly wider spread than the signal.  
We can, however, estimate this background contribution by analyzing the $\dm$ 
distribution in which signal events have a Gaussian
shape and background events have a threshold function shape as shown in Fig.~\ref{fg:fit_dm_scf3}.

The signal yields the of $\bm \to \pldstzp$ modes are extracted from 
a two-dimensional extended unbinned maximum likelihood fit with the likelihood defined as 
\begin{equation}
L = \frac{e^{-\sum_{j}N_j}}{ N!}\prod_{i=1}^{N}(\sum_{j}N_jP_i^j),
\end{equation}
where $N$ is the total number of candidate events, $N_j $ denotes the number of corresponding category events and $P_i^j$ represents the corresponding
two-dimensional PDF in $\mb$ and $\de$; $i$ denotes the $i$-th event,
and $j$ indicates the index of different event categories in the fit.
Thus, $j$ could either indicate signal or combinatorial background for the $\pld$ case and includes one more category (CF) for the $\pldst$ case. 
We use a Gaussian function to represent the $\mb$ signal and a double Gaussian function for the 
 $\de$ signal with parameters determined using MC simulations.
Combinatorial background is described by an ARGUS function~\cite{Argus} and 
a second-order Chebyshev polynomial in the $\mb$ and $\de$ distributions, respectively.

Since it is difficult to separate the $\bm \to \pldst$ signal and CF events in the fit, 
we estimate the number of CF events in the $\dm$ signal region 
($0.139$ GeV/$c^2 < \dm <$ 0.145 GeV$/c^2$) from the fitted CF yield in the $\dm$ sideband 
region ($0.15$ GeV/$c^2 < \dm < 0.17$ GeV$/c^2$). 
The ratio of the area of the CF in the $\dm$ signal region to that in the sideband region is $26.0\pm0.9\%$, which is determined from MC samples of $\bm \to \pldst$ (Fig.~1) and $\overline{\bz} \to \pldstp$.
The PDF used for CF events is a product of a Gaussian-like smoothed histogram for $\mb$ and a 
double Gaussian function for $\de$ with parameters determined using MC simulations.   
We fix the number of CF events in the $\mb-\de$ fit to determine the 
signal yield within the $\dm$ signal region.

\begin{figure}[htb]
 \hskip  2.35 cm  {\bf{(a)}} \hskip  3.15 cm  {\bf{(b)}}
 \vskip -1.2cm
 \includegraphics[width=0.45\textwidth]{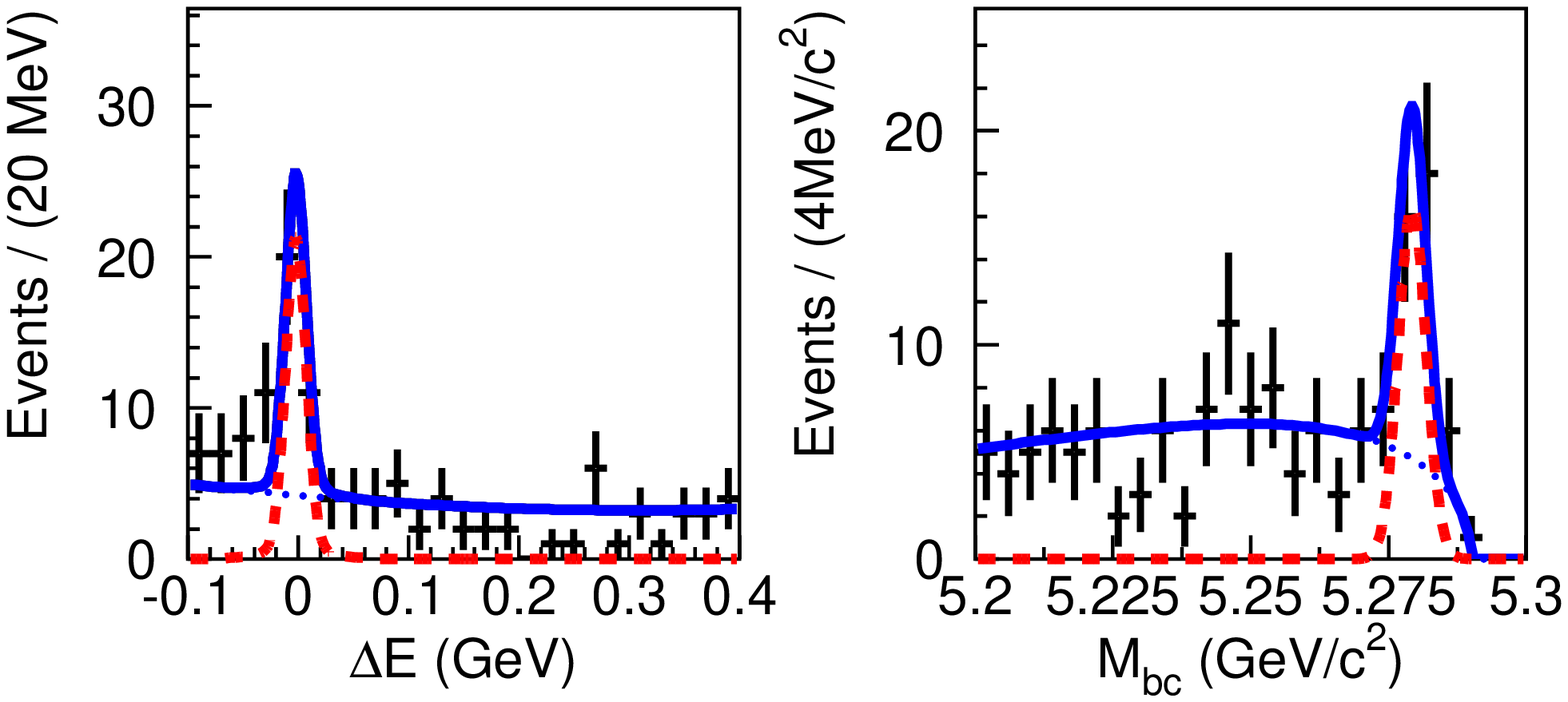}\\
 \hskip  2.35 cm  {\bf{(c)}} \hskip  3.15 cm  {\bf{(d)}}
  \vskip -1.2cm
\includegraphics[width=0.45\textwidth]{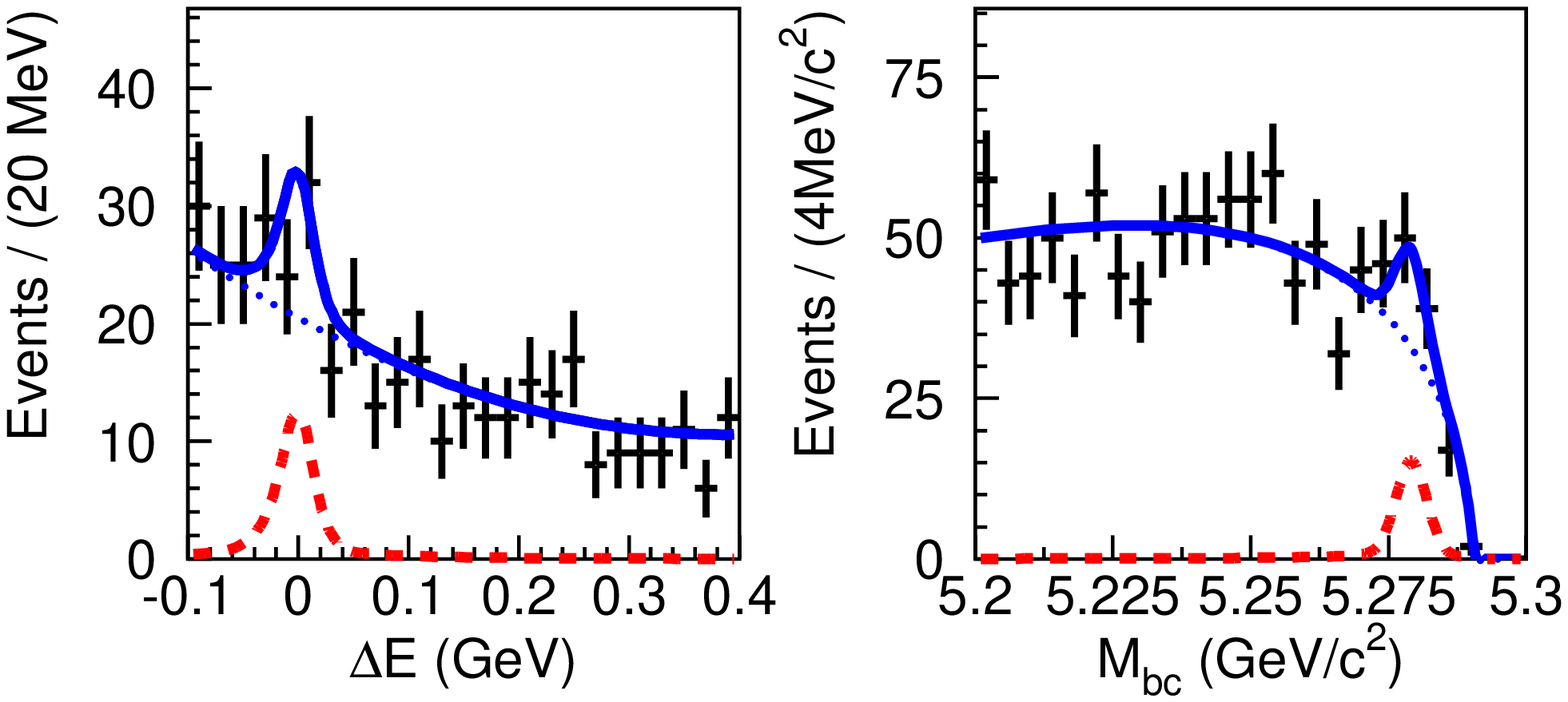}\\
\caption{ 
Distributions of $\de$ (a, c) for $\mb > 5.27$ GeV/$c^2$ and of $\mb$ (b, d) for $|\de| <0.05 $ GeV; 
the top row is the fit result for $\bm \to \pld$, $D^0 \to K^-\pi^+ $ (a, b) and the bottom row for $\bm \to \pld$, $D^0 \to K^-\pi^+ \pi^0$ (c, d). 
The points with error bars are data; the solid
curve shows the fit; the dashed curve represents the signal, 
and the dotted curve indicates continuum background.}
\label{fg:data_fit1}
\end{figure}

Figure \ref{fg:data_fit1} shows the result of the fit for $\bm \to \pld$.
The fitted signal yields in the data sample are $\pldtwoyd$ and $\pldthryd$ events with statistical significances of 7.6 and 3.6 standard deviations ($\sigma$) for $\bm \to \pld, ~D^0 \to K^-\pi^+$ and $D^0 \to K^-\pi^+\pi^0$, respectively.
The significance is defined as $\sqrt{-2{\rm ln}(L_0/L_{\rm max})}$,
where $L_0$ and $L_{\rm max}$ are the likelihood values returned by the fit with a signal yield fixed to zero and  the nominal fit, respectively. 
The branching fractions are calculated using the formula
\begin{eqnarray}
\nonumber \mathcal{B} =\frac{N_{signal}}{\epsilon \times f \times N_{B\overline{B}} },
\end{eqnarray}
where $N_{signal}$, $N_{B\overline{B}}$, $\epsilon$, and $f$ are the fitted number of signal events, the number of $B\overline{B}$ pairs, the reconstruction efficiency, and the relevant sub-decay branching fractions:
$\mathcal{B}(\Lambda \to p\pi^-) ~= 63.9 \pm 0.5 \%$,
$\mathcal{B}(D^0 \to K^-\pi^+)  ~= 3.89 \pm 0.05\% $,
$\mathcal{B}(D^0 \to K^-\pi^+\pi^0)  ~= 13.9 \pm 0.5\% $, and
$\mathcal{B}(D^{*0} \to D^0\pi^0) ~= 61.9 \pm 2.9\%$~\cite{3}. 
We assume that charged and neutral $B\overline{B}$ pairs are equally produced at the $\Upsilon(4S)$.

\begin{figure}[htb]
\begin{sideways}   {~~~~~~~~~~~~~~~~~~~~$d\mathcal{B} / dM_{\plbar}$ $(10^{-6})$ / ( 0.2 GeV /$c^2)$ }  \end{sideways}
 \hskip  -0.5 cm 
{\includegraphics[width=0.35\textwidth]{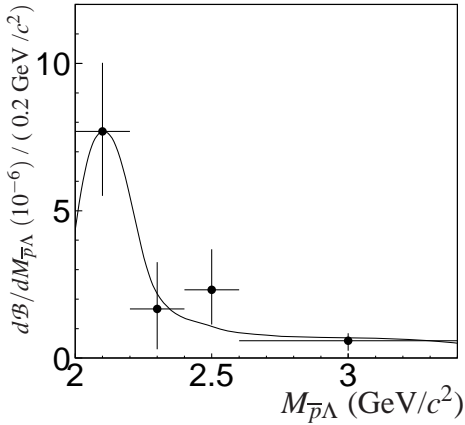}}\\
\vskip -0.9cm  
\hskip  3.5 cm  
 {\large $M_{\plbar} $ (GeV/$c^2$)}
\caption{ Differential branching fraction ($d\mathcal{B} / dM_{\plbar}$) as a function of the $\plbar$ mass for $\bm \to \pld$.
Note that the last bin with the central value of 3 GeV/$c^2$ has a bin width of 0.8 GeV/$c^2$. The solid curve is a fit with a threshold function.
}
\label{fg:pld0_mpl}
\end{figure}

To investigate the threshold enhancement feature, we
determine the differential branching fractions in bins of $M_{\plbar}$; 
the results obtained from the weighted averages of the fits to $\bm \to \pld$, $D^0 \to K^-\pi^+$ and $D^0 \to K^-\pi^+\pi^0$ separately are shown in Fig.~\ref{fg:pld0_mpl} where an enhancement near threshold is evident.
We fit the $\plbar$ mass spectrum with a threshold function and then reweight MC events to match the fitted threshold
function in order to obtain a proper estimate of the reconstruction efficiency 
for signal events. 
The observed branching fractions are 
$ \pldtwobf$ for $\bm \to \pld$ with $D^0 \to K^-\pi^+$ and
$ \pldthrbf$ for $\bm \to \pld$ with $D^0 \to K^-\pi^+\pi^0$. 
The weighted average of the branching fractions is $\pldtotbf$ with a significance of 8.1$\sigma$, where the
systematic uncertainties (described below) on the signal yield are also included in the significance evaluation.

\begin{figure}[htb]
 \vskip  0.0cm
 \hskip  2.cm  {\bf{(a)}} \hskip  3.15 cm  {\bf{(b)}}
 \vskip -1.2cm
\includegraphics[width=0.43\textwidth]{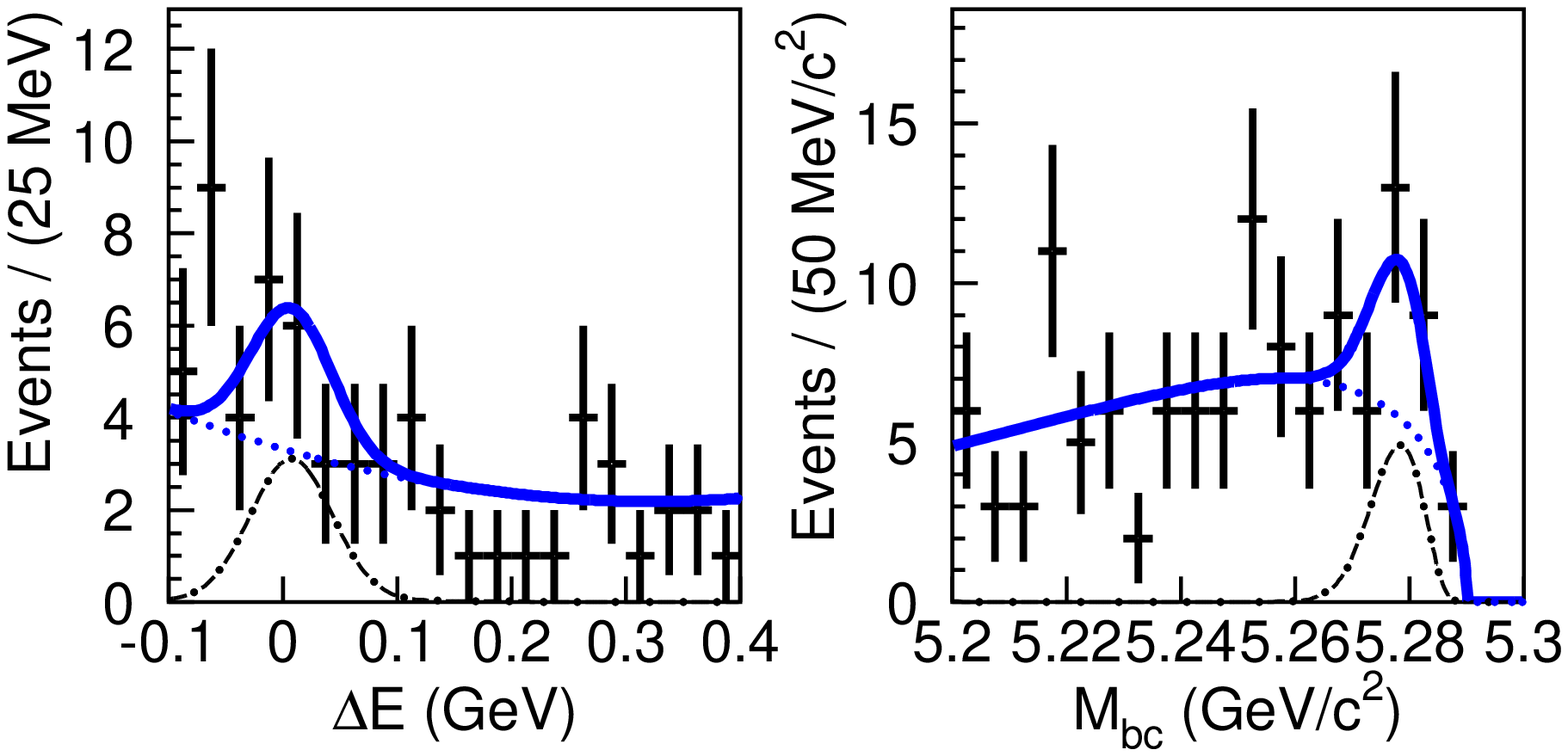}\\
 \hskip  2. cm  {\bf{(c)}} \hskip  3.15 cm  {\bf{(d)}}
  \vskip -1.2cm
\includegraphics[width=0.43\textwidth]{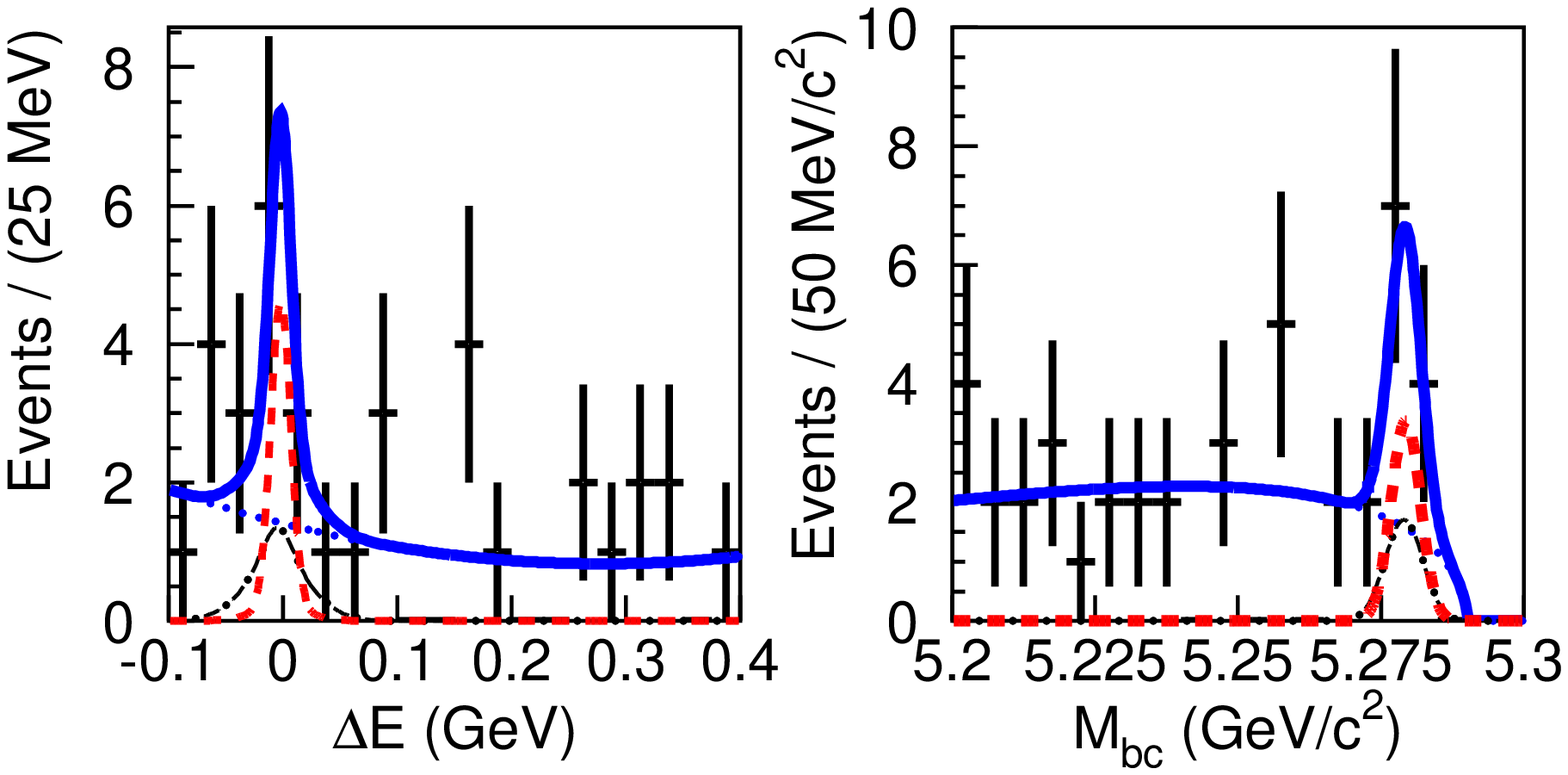}
\caption{
Distributions of $\de$ (a, c) for $\mb > 5.27$ GeV/$c^2$ and of $\mb$ (b,d) for $|\de| <0.05 $ GeV; 
the top row is the fit result for $\bm \to \pldst$ in the $\dm$ sideband region (a,b) and the bottom row for $\bm \to \pldst$ in the $\dm$ signal region (c, d). 
The points with error bars are data; the solid curve shows the result of the fit; 
the dot-dashed and dotted curve indicates the CF and continuum background; 
the dashed curve represents the signal.}
\label{fg:data_fit2}
\end{figure}

The fit results for $\bm \to \pldst$ in the $\dm$ sideband region are shown in Fig.~\ref{fg:data_fit2} (a, b). 
The number of CF events in the sideband  is $11.6 \pm 5.4$, which is used to estimate the number of CF events in the $\dm$ signal region, $3.0 \pm 1.4$, after scaling by the 
 area ratio of CF (26.0$\pm$ 0.9\%). 
We then fix the normalization of the CF component 
in the fit to the $\dm$ signal region [fit results are shown in Fig.~\ref{fg:data_fit2} (c, d)], and 
obtain a signal yield of $\pldstzyd$ with a statistical significance of 2.2$\sigma$.
Assuming $\bm \to \pldst$ and $\bm \to \pld$ have the same $\plbar$ spectrum,
we determine $\mathcal{B}(\bm \to \pldst)$ to be $\pldstzbf$. 
In the absence of a statistically compelling signal yield,
 we set an upper limit $\mathcal{B}(\bm \to \pldst) < 4.8 \times 10^{-5} $ at the 90\% confidence level
using the Feldman-Cousins method~\cite{FC, pole}. 
The information used to obtain the upper limit includes the number of events in the signal region (13) and $8.1  \pm 1.4$ background events.
Here, the background that is integrated in the signal region, consists of 5.3$\pm$0.5 continuum events and 2.9$\pm$1.3 CF events from the fit to the $\dm$ sideband. 
The 11.7\% additive systematic uncertainty due to the selection criteria is included in the determination  of the upper limit on the branching fraction.\\

\begin{table}
\caption{ Summary of the results: event yield, significance, efficiency, and branching
fraction.}
\tabcolsep= 6pt
\renewcommand{\arraystretch}{1.5}
\begin{center}
\begin{tabular}{clcccc}
\hline\hline
 Mode        & $N_{signal}$ &	$\mathcal{S}$        &   $\epsilon$(\%) &	    $\mathcal{B}( 10^{-5})$  \\
\hline\hline 
$\overline{p} \Lambda D^0_{K^-\pi^+}$ 	         & $\pldtwoyd$    & 7.4  &	 11.7 & 	 $\pldtwobfx$\\
$\overline{p} \Lambda D^0_{K^-\pi^+ \pi^0}$     & $\pldthryd$    & 3.4  &	 4.0  &   $\pldthrbfx$	\\    
\hline
$\bm \to \pld$	 						    	 &               & 8.1  &	 		   &   $\pldtotbfx$	\\ 
\hline\hline 
$\bm \to \pldst$      			 & $\pldstzyd$    & 2.1  &	 2.8 & 	 $\pldstzbfx$\\
\hline\hline
\end{tabular}
\end{center}
\label{data_yields}
\end{table}

Systematic uncertainties are estimated using high-statistics control samples.
A track reconstruction efficiency uncertainty of 1.2\% is assigned for each track.
For the proton identification efficiency uncertainty, we use a $\Lambda \to p \pi^-$ sample, and for
$K-\pi$ identification uncertainty we use a sample of kinematically identified  $D^{*+} \to D^0\pi^+$,
 $D^0 \to K^-\pi^+$ decays. 
The average efficiency discrepancy due to hadron identification differences 
between data and MC simulations has been corrected for the final branching fraction measurements. 
The corrections due to the  hadron identification are 10.7\% and 10.6\% for $\bm \to \pld$ and $\bm \to \pldst$, respectively.
The uncertainties associated with the  hadron identification corrections are 4.2\% for 
two protons (one from $\Lambda$ decay), 0.5\% for a charged pion, and 1.0\% for a charged kaon.

The $\pi^0$ selection uncertainty is found to be 5.0\% by comparing the ratios of efficiencies between $D^0 \to K^-\pi^+$ and $D^0 \to K^-\pi^+\pi^0$ for data and MC samples.
In the $\Lambda$ reconstruction, we find an uncertainty of $4.1\%$ from the differences between data and MC for the efficiencies of tracks displaced from the interaction point, the $\Lambda$ proper time distributions, and the $\Lambda$ mass spectrum.
The uncertainty due to the $\mathcal R$ selection for $\bm \to \pld$, $D^0 \to K^-\pi^+$ is estimated 
from the control sample $\bm \to {D}^0 \pi^-$, $D^0 \to K^-\pi^+\pi^-\pi^+$ and is determined to be 1.3\%.
The $\mathcal R$ related uncertainty for $\bm \to \pld$, $D^0 \to K^-\pi^+ \pi^0$ is 3.0\% estimated 
from $\bm \to {D}^0 \pi^-$, $D^0 \to K^-\pi^+\pi^0$.
The uncertainties due to the $D^0$ mass selection for $D^0 \to K^-\pi^+$ and $D^0 \to K^-\pi^+\pi^0$ are 1.9\% and 1.6\%, respectively.

The dominant systematic uncertainty for $\bm \to \pld$ is due to the modeling of PDFs, estimated by including a $B \to \pld\pi$ or a nonresonant $\bm \to \overline{p} \Lambda K^-\pi^+ (K^-\pi^+ \pi^0)$ component in the fit, 
modifying the efficiency after changing the signal $\mpl$ distribution, and varying the parameters of the signal and background PDFs by one standard deviation using MC samples.
The modeling uncertainties are 7.5\% and 12.9\% for $\bm \to \pld$ with $D^0 \to K^-\pi^+$ and $D^0 \to K^-\pi^+ \pi^0$, respectively. 
The overall modeling uncertainty for $\bm \to \pldst$ of 28.6\% is obtained from two kinds of PDF modifications.
The parameters of the fixed CF component are varied by their $\pm 1\sigma$ statistical uncertainties, which were obtained from the fit to the $\dm$ sideband region.
We also include an additional PDF for the 
combinatorial background based on the \textsc{Pythia}~\cite{Pythia} $b$ quark fragmentation process, e.g.,
$\bm \to \overline{p} \Lambda {D}^{0}$, $\bp \to \overline{p} \Delta^{++} {D}^{*0}$, $\bm \to \overline{p} \Delta^{0} {D}^{*0}$, $\bm \to \overline{p} \Sigma^0 {D}^{*0}$, etc.

The systematic uncertainties from the sub-decay branching fractions are calculated from the corresponding 
branching uncertainties in~\cite{3}; they are 1.5\% (3.7\%) and 6.0\% 
for $\bm \to \pld,~D^0 \to K^-\pi^+$ ($D^0 \to K^-\pi^+\pi^0$) and $\bm \to \pldst$, respectively.
The uncertainty in the number of $B\overline{B}$ pairs is 1.4\%.
The total systematic uncertainties are 11.6\% (17.1\%) and 30.9\% for $\bm \to \pld$ with $D^0 \to K^-\pi^+$
 ($D^0 \to K^-\pi^+ \pi^0$) and $\bm \to \pldst$, respectively.
The final results are listed in Table~\ref{data_yields}, where the significance values are
modified and include the systematic uncertainty related to PDF modeling.

In summary, using a sample of $657 \times 10^6~B\overline{B}$ events, we report the first 
observation of $\bm \to \pld$ with a 
branching fraction of $\pldtotbf$ and a significance of 8.1$\sigma$.
No significant signal is found for $\bm \to \pldst$ and the corresponding upper limit is $4.8 \times 10^{-5}$ at the 90\% confidence level.
We also observe a $\plbar$ enhancement near
threshold for $\bm \to \pld$, which is similar to a common feature found in charmless three-body 
baryonic $B$ decays~\cite{3}. 
The measured $\bm \to \pld$ branching fraction agrees with the theoretical prediction 
of $(1.14 \pm 0.26) \times 10 ^{-5} $~\cite{5}. This indicates that the generalized factorization approach with parameters determined from experimental data gives reasonable estimates for $b \to c$ decays. 
This information can be helpful for future theoretical studies of the angular distribution puzzle in the penguin-dominated processes, $B^- \to p\overline{p}K^{-}$ and $\bz \to \plpi$~\cite{5}.
The measured branching fraction for $\bm \to \pld$ can also be used to tune the parameters in the event generator, e.g., 
\textsc{Pythia}, for fragmentation processes involving $b$ quarks. 
Although the current statistics for $\bm \to \pld$ are still too low to perform
an angular analysis of the baryon-antibaryon system, the proposed super flavor factories~\cite{SFF1,SFF2}
 offer promising venues for such studies. 
\\

We thank the KEKB group for the excellent operation of the
accelerator, the KEK cryogenics group for the efficient
operation of the solenoid, and the KEK computer group and
the National Institute of Informatics for valuable computing
and SINET4 network support.  We acknowledge support from
the Ministry of Education, Culture, Sports, Science, and
Technology (MEXT) of Japan, the Japan Society for the 
Promotion of Science (JSPS), and the Tau-Lepton Physics 
Research Center of Nagoya University; 
the Australian Research Council and the Australian 
Department of Industry, Innovation, Science and Research;
the National Natural Science Foundation of China under
contract No.~10575109, 10775142, 10875115 and 10825524; 
the Ministry of Education, Youth and Sports of the Czech 
Republic under contract No.~LA10033 and MSM0021620859;
the Department of Science and Technology of India; 
the BK21 and WCU program of the Ministry Education Science and
Technology, National Research Foundation of Korea,
and NSDC of the Korea Institute of Science and Technology Information;
the Polish Ministry of Science and Higher Education;
the Ministry of Education and Science of the Russian
Federation and the Russian Federal Agency for Atomic Energy;
the Slovenian Research Agency;  the Swiss
National Science Foundation; the National Science Council
and the Ministry of Education of Taiwan; and the U.S.\
Department of Energy.
This work is supported by a Grant-in-Aid from MEXT for 
Science Research in a Priority Area (``New Development of 
Flavor Physics''), and from JSPS for Creative Scientific 
Research (``Evolution of Tau-lepton Physics'').


\begin{thebibliography}{99}
\bibitem{10} Unless otherwise stated, charge-conjugate decays are 
implicitly assumed throughout the paper.
\end{thebibliography}
\end{document}